\documentclass[%
reprint,
superscriptaddress,
amsmath,amssymb,
aps,
]{revtex4-2}
\usepackage{mathtools}
\usepackage{textgreek}
\usepackage{tabularx}
\usepackage{subfigure}
\usepackage{multirow}
\usepackage{graphicx}
\usepackage{dcolumn}
\usepackage{bm}
\usepackage{xcolor}
\usepackage{float}
\usepackage[utf8]{inputenc}
\usepackage[english]{babel}
\UseRawInputEncoding
\usepackage{subfigure}

\usepackage{placeins}

\usepackage[subsectionbib]{bibunits}
\defaultbibliographystyle{apsrev4-2}
\defaultbibliography{references}

\usepackage{hyperref}

\begin{document}
\preprint{APS/123-QED}

\title{Methods to achieve near-millisecond energy relaxation and dephasing times for a superconducting transmon qubit}

\author{Mikko Tuokkola}
\email{Corresponding author email: mikko.tuokkola@aalto.fi}
\author{Yoshiki Sunada}
\author{Heidi Kivijärvi}
\author{Jonatan Albanese}
\affiliation{%
 QCD Labs, QTF Centre of Excellence, Department of Applied Physics,\\
 Aalto University, P.O. Box 13500, FIN-00076 Aalto, Finland 
}%
\author{Leif Grönberg}
\author{Jukka-Pekka Kaikkonen}
\author{Visa Vesterinen}
\author{Joonas Govenius}
\affiliation{%
VTT Technical Research Centre of Finland Ltd. \& QTF Centre of Excellence, P.O. Box 1000,
02044 VTT, Finland
}%
\author{Mikko Möttönen}
\email{Corresponding author email: mikko.mottonen@aalto.fi}
\affiliation{%
 QCD Labs, QTF Centre of Excellence, Department of Applied Physics,\\
 Aalto University, P.O. Box 13500, FIN-00076 Aalto, Finland 
}%
\affiliation{%
VTT Technical Research Centre of Finland Ltd. \& QTF Centre of Excellence, P.O. Box 1000,
02044 VTT, Finland
}%

\date{\today}

\begin{abstract}

Superconducting qubits are one of the most promising physical systems for implementing quantum computers. However, executing quantum algorithms of practical computational advantage requires further improvements in the fidelities of qubit operations, which are currently limited by the energy relaxation and dephasing times of the qubits. Here, we report our measurement results of a high-coherence transmon qubit with energy relaxation and echo dephasing times surpassing those in the existing literature. We measure a qubit frequency of 2.9~GHz, an energy relaxation time $T_1$ with a median of $425$~\textmu{s} and a maximum of $(666 \pm 33)$~\textmu{s}, and an echo dephasing time $T_2^\mathrm{echo}$ with a median of $541$~\textmu{s} and a maximum of $(1057 \pm 138)$~\textmu{s}. We report in detail our design, fabrication process, and measurement setup to facilitate the reproduction and wide adoption of high-coherence transmon qubits in the academia and industry. 

\end{abstract}

\maketitle

\begin{bibunit}

\section{Introduction} \label{sec:Introduction}

In the past two decades, the rapid development of superconducting qubits has rendered them one of the most promising candidates for realizing large-scale quantum computers~\cite{kjaergaard_superconducting_2020, krantz_quantum_2019, mamgain_review_2023, huang_superconducting_2020, ezratty_perspective_2023, acharya_quantum_2024}. 
The transmon qubit, proposed in 2007, has become the most widely used superconducting qubit due to its simplicity and performance~\cite{koch_charge-insensitive_2007,blais_circuit_2021}. 
It consists of a Josephson junction and a shunt capacitor which reduces the sensitivity of the qubit to charge noise but keeps the qubit sufficiently anharmonic for realizing fast, high-fidelity gate operations.

\begin{figure}[tb!]
\centering
\includegraphics{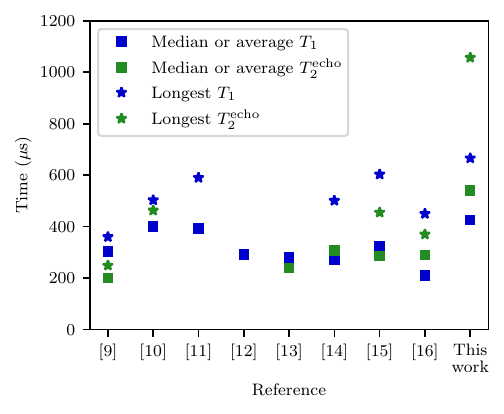}
\caption{\textbf{Summary of the longest measured energy relaxation and echo dephasing times in the existing literature}. Squares represent the median or average values, and stars represent the longest values reported in the references which are chronologically ordered.
\label{fig:coherence_times_literature_image} }
\end{figure}

Short coherence times have historically been a disadvantage of superconducting qubits, but the previous decade has shown continuous improvements on this challenge, leading to increased fidelities of qubit operations. The longest reported energy relaxation times $T_1$ of transmons have approached but not surpassed 400-\textmu s median value as shown in Fig.~\ref{fig:coherence_times_literature_image} and Extended Data Table~\ref{tab:previous-works} based on Refs.~\cite{place_new_2021, wang_towards_2022, gordon_environmental_2022, deng_titanium_2023,sivak_real-time_2023,biznarova_mitigation_2023, bal_systematic_2024,kono_mechanically_2024}. Recently, there have also been significant improvements in transmon echo dephasing times $T_2^\mathrm{echo}$ with the longest reported average extending to 307~\textmu s~\cite{wang_towards_2022, sivak_real-time_2023,biznarova_mitigation_2023, kono_mechanically_2024}. Transmon qubits with coherence times above 100~\textmu s have also been reported, for example, in Refs.~ \cite{dunsworth_characterization_2017, wang_cavity_2019, nersisyan_manufacturing_2019, wei_verifying_2020, moskalev_optimization_2023, pishchimova_improving_2023}. With other types of superconducting qubits such as 0-$\pi$ qubits and fluxonium qubits, energy relaxation times above a millisecond have been observed, but these qubits have either had short dephasing times or low qubit frequencies~\cite{pop_coherent_2014,gyenis_experimental_2021,somoroff_millisecond_2023}.

Here, we report our latest results on high-coherence transmon qubits with improved energy relaxation and echo dephasing times. Importantly, we describe in detail our design, fabrication process, and measurement setup with the intention to promote the reproduction of our result by other academic and industrial groups.

\section{Sample and its fabrication} \label{sec:Fabrication process}

\begin{figure}[b!]
\centering
\includegraphics[width=0.45\textwidth]{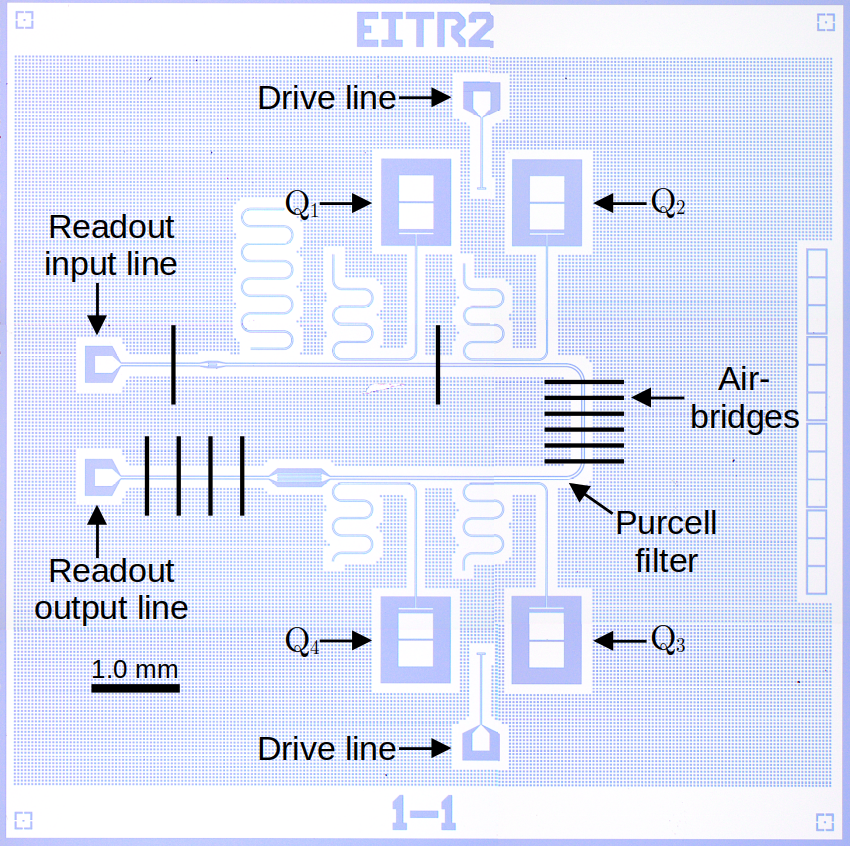}
\caption{\label{fig:sample_image} Microscope image of a sample which is identical to the one measured in this work. The sample contains four transmon qubits ($\mathrm{Q}_1$--$\mathrm{Q}_4$), their readout resonators, and a shared Purcell filter. Qubits $\mathrm{Q}_1$ and $\mathrm{Q}_2$ share a single drive line, and $\mathrm{Q}_3$ and $\mathrm{Q}_4$ share another one. Air-bridges are made by wire bonds using aluminum wires.}
\end{figure}

Figure~\ref{fig:sample_image} shows a microscope image of a sample which is identical to the one measured in this work. It contains four transmon qubits $\mathrm{Q}_1$--$\mathrm{Q}_4$, each of which couples to a coplanar waveguide resonator for readout. Qubits Q$_1$ and Q$_3$ are flux-tunable by the use of a superconducting quantum interference device (SQUID), whereas Q$_2$ and Q$_4$ are fixed-frequency qubits. The four qubit--resonator pairs are identical except for the size of the Josephson junction and the length of the resonator. The readout resonators couple to a shared Purcell filter, which reduces the energy relaxation of the qubits into the readout lines~\cite{jeffrey_fast_2014,sete_quantum_2015}. The layout design file of the sample is available in Ref.~\cite{tuokkola_data_2024}. For more details of the sample, see Extended Data Figs.~\ref{fig:gds_images} and~\ref{fig:sem_images}

This section provides a detailed description of our sample fabrication process, which is an adaptation of the recipe described in Ref.~\cite{kono_mechanically_2024} to the chemicals and equipment available to us. Notable differences include the choice of CF$_4$ as the processing gas for Nb etching, which improves the reproducibility of the etch, and the pre-dicing of the sample between the Al evaporation and lift-off, which minimizes the exposure of the sample to the ambient atmosphere. The pieces of equipment used in our process are listed in Extended Data Table~\ref{tab:fab equipments}.

\subsubsection{Substrate and niobium patterning}
The sample is fabricated using a 675-\textmu m-thick 6-inch (100)-oriented high-resistivity ($>$10~k$\Omega$cm) intrinsic-silicon wafer sourced from Siegert Wafer. The pre-cleaning of the wafer follows Ref.~\cite{burnett_noise_2018} and begins with an RCA solvent clean. The wafer then undergoes a dip in dilute hydrofluoric (HF) acid (1:100) for 1~min. After being rapidly transferred to the sputtering tool to minimize exposure to the ambient atmosphere, the wafer is baked at 300~$^\circ$C under vacuum. The sputtering of a 200-nm Nb film is carried out near room temperature and at 2600-W power. The sputter target has a purity of 99.998\%. 
After the sputtering, the wafer is coated with a protective layer of AZ~5214E photoresist and diced into 25~mm~$\times$~30~mm rectangular coupons with a dicing saw.

The resonators, coplanar waveguides, ground plane, and transmon capacitors are patterned as described below. The selected coupon is sonicated in acetone and isopropyl alcohol (IPA) for 3~min each to remove the protective resist layer and dried with a nitrogen gun. The sample is then dehydrated on a hotplate at 110~$^{\circ}$C for 1~min, spin coated with AZ 5214E photoresist at 4000~rpm, and baked at 110~$^{\circ}$C for 1~min to achieve a coating thickness of 1.4~\textmu{m}. The photoresist is exposed using a maskless aligner with a laser wavelength of 405~nm and a dose of 130 mJ$/$cm$^2$. Subsequently, the photoresist is developed in AZ~726MIF for 1.5~min and rinsed in deionized water (DIW) for 1.5~min.

The Nb film is patterned in a plasma processing system by a chemical dry-etching process, clearing the areas revealed by the resist development step. Immediately before the process, the empty plasma chamber is cleaned with a combination of CF$_4$ and O$_2$ gases for a total of 5~min. We then load the sample into the chamber and apply oxygen plasma ashing for 10~s at 100~mTorr chamber pressure, 40~sccm gas flow, and 150~W rf source power to remove the post-development residue of photoresist on the sample surface. Then, we pump out the oxygen and introduce CF$_4$ at 50-mTorr chamber pressure, 20-sccm flow, and 30-W source power. To ensure that the Nb film is fully etched, we carry out an additional etch after unloading, visually inspecting, and re-loading the sample. The etch rate of Si is significantly lower than that of Nb, which helps us to achieve a convenient control of the etch depth. The typical total etching time of Nb film is around 10~min. After the etching process, we remove some of the residual chemicals by applying another oxygen plasma ashing for 2~min without breaking the vacuum.

The photoresist is removed by immersing the sample in an $N$-methylpyrrolidone (NMP)-based solvent Remover PG at 80~$^{\circ}$C overnight and sonicating it in the same Remover PG, acetone, and IPA for 3~min each. Subsequently, the sample is dried with a nitrogen gun and plasma-ashed again for 2~min.
Using a profilometer, we measure the etch depth to be 250~nm, which implies that the Si substrate is etched by 50~nm.

\subsubsection{Electron-beam lithography}
In order to remove the oxide layers on the Nb and Si surfaces, we immerse the sample in dilute HF acid (0.5\%) for 10~min and rinse it in DIW for 5~min~\cite{verjauw_investigation_2021}. The sample is then carried to a spin coater while immersed in fresh DIW to minimize its exposure to the ambient atmosphere. After spin-drying the sample, we immediately spin coat the methyl methacrylate (MMA) EL11 copolymer resist (11\% solid content in ethyl lactate) at 4000~rpm, bake it at 180~$^{\circ}$C for 5~min, and cool it for 3~min. Then we coat the sample with polymethyl methacrylate (PMMA) 950 A4 resist (950,000 molecular weight, 4\% solid content in anisole) at 1000~rpm and bake it at 180~$^{\circ}$C for 5~min. This creates a two-layer resist stack with approximately 500~nm of MMA and 400~nm of PMMA.

The resist mask for Manhattan-style Josephson junctions is patterned onto the resist stack using an electron-beam writer with 100-kV acceleration voltage, 300-\textmu m aperture, and 0.5-nA beam current. We use a dose of 1000~\textmu C/cm$^2$ to define the junction structure in both the PMMA layer and the MMA layer. In addition, we use a dose of 400~\textmu C/cm$^2$ to define an undercut at each end of the line-like structures for the junctions. The undercut serves to separate the aluminum junction evaporated onto the Si substrate from the MMA side walls. We use a mask pattern very similar  to that of Ref.~\cite{kono_mechanically_2024} to define the Josephson junctions. The electron-beam resist is developed by immersing the sample in methyl isobutyl ketone (MIBK):IPA (1:3) solvent for 5~min, rinsing it in IPA for 1~min, and drying it with a nitrogen gun.

\subsubsection{Junction deposition}
The Josephson junctions are deposited using an ultra-high-vacuum electron-beam evaporator with separate load-lock, oxidation, and evaporation chambers. After loading the sample, we pump the system for 14~h to reach a load-lock chamber pressure below $10^{-7}$~mbar. Subsequently, we carry out ozone ashing at 10~mbar for 1~min to remove a thin layer of resist residues. 

After the ozone cleaning, a high vacuum ($<$10$^{-7}$~mbar) is pumped again to the load-lock chamber, and the sample is transferred to the oxidation/evaporation chamber. To fabricate the junctions, we evaporate Al at the rate of 0.2~nm/s at tilt and in-plane rotation angles specific for each of the four Al line strips as discussed below. Prior to each Al evaporation step, the oxidation and evaporation chambers are getter-pumped by evaporating Ti with a closed shutter at the rate of 0.1~nm/s for 2~min and waiting for the chamber pressure to decrease below $5\times 10^{-8}$~mbar.

For the bottom layer of the junctions, we deposit 40~nm of Al at $\theta=45^{\circ}$ tilt and $\phi=-45^{\circ}$ planetary angle. Then we create the insulating aluminum oxide layer by static oxidation at 1.2~mbar for 5~min. During the oxidation, the aluminum source is protected from the oxidation by a valve that blocks the oxygen flow.  After the oxidation, we deposit the second Al layer in two steps with $\theta=45^{\circ}$ tilt and two planetary angles: first 30~nm at $\phi=45^{\circ}$ and then 30~nm at $\phi=-135^{\circ}$. This ensures that both sides of the oxidized bottom strip are covered by the second Al layer. 

In order to ensure a galvanic contact between the Josephson junctions and the Nb capacitor pads, we transfer the sample to the load-lock chamber and carry out argon milling with $\theta=45^{\circ}$ tilt at two planetary angles $\phi=\pm 90^{\circ}$ for 2~min each at 10-sccm gas flow, 400-V beam voltage, 60-mA beam current, and 80-V acceleration voltage. This removes any Nb surface oxide that has grown in the areas exposed after the electron-beam resist development~\cite{osman_simplified_2021, bilmes_-situ_2021}.
After the argon milling, the sample is transferred back to the oxidation/evaporation chamber, where we deposit the connecting leads between the junction and the Nb pads in three steps with 30/60/60-nm thicknesses, $\theta=45^{\circ}$ tilt and $\phi=180^{\circ}/0^{\circ}/180^{\circ}$ planetary angles.

As the final step, the sample is transferred back to the load-lock chamber and oxidized at 20~mbar for 10~min to create a clean oxide layer on top of the junctions before exposing the sample to the ambient atmosphere.

\subsubsection{Dicing and liftoff}
After the Josephson junction deposition, we coat the sample with a protective layer of AZ~5214E photoresist and pre-dice the sample with a dicing saw, cutting two-thirds deep of the substrate thickness. Before attaching the sample to the dicing tape, an ionizer fan is applied to the tape to reduce the possibility of electrostatic-discharge damage on the sample. By pre-dicing the sample prior to the liftoff, we can skip one of the resist removal steps carried out in Ref.~\cite{kono_mechanically_2024}.

The pre-diced sample is immersed in Remover PG at 80~$^{\circ}$C for 3~h, after which large Al flakes can be removed from the beaker using a pipette. The sample is sonicated in the same Remover PG, acetone, and IPA for 3~min each. From IPA, the sample is quickly dried with a nitrogen gun and immediately transferred to a vapor prime oven that applies a monolayer of hexamethyldisilazane (HMDS) on the sample surface, which may help to slow down the post-fabrication oxidation of the Nb and Si surfaces. The oven also has the effect of annealing the sample at 150$^\circ$C for a total of 20~min under vacuum and nitrogen environments.

The critical currents of the fabricated junctions are estimated by measuring their room-temperature resistances using a probe station. The critical currents for the sample measured in this work are significantly smaller than we had targeted, which led to low $E_J/E_C$ ratios, as low as 20 for qubit Q$_2$. The sample is then manually cleaved into separate chips utilizing the cuts established with the dicing saw, and selected chips are taken for further characterization.

\section{Setup for qubit measurements} \label{sec:Experimental setup}

\begin{figure}[tb!]
\includegraphics{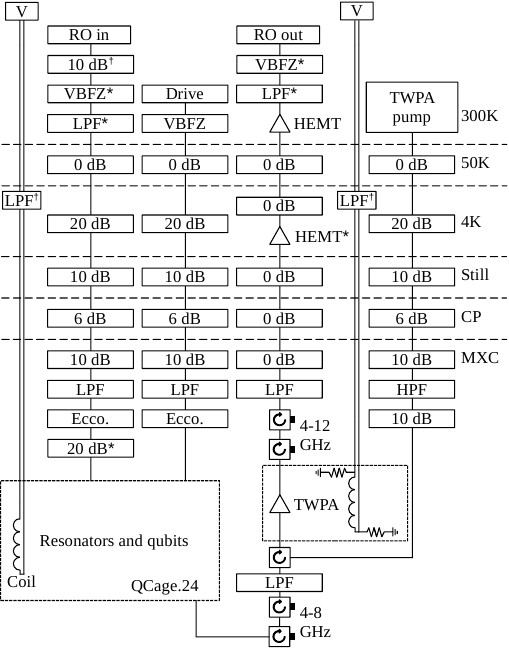} 
\caption{\textbf{Schematic of the experimental setup for qubit characterization and measurement.} See Extended Data Table~\ref{tab:measurement equipments} for a description of each component. For the second cooldown, the $10$-dB attenuator in the drive line at the mixing-chamber (MXC) plate is removed, and the first $10$-dB attenuator in the TWPA pump line at the MXC plate is replaced with an Eccosorb filter. \label{fig:schematic_of_setup}}
\end{figure}

Our experimental setup used for qubit characterization and measurements is presented in Fig.~\ref{fig:schematic_of_setup}, with detailed information on measurement equipment and components provided in Extended Data Table~\ref{tab:measurement equipments}. The sample is wire-bonded to the printed circuit board (PCB) of a QCage.24 sample holder and placed inside QCage Magnetic Shielding, both of which are supplied by QDevil (under Quantum Machines). The chip is suspended by four corners inside a cavity and clamped down by the PCB. The assembly is placed inside a light-tight superconducting aluminium enclosure to reject stray microwave and infrared photons before being mounted inside the magnetic shield. The sample is exposed to ambient atmosphere at room temperature for a total of 7 days. It is then cooled down to approximately 10~mK in a Bluefors dilution refrigerator equipped with a tin-plated copper shield at 10~mK and an Amumetal magnetic shield just inside the outer vacuum chamber~(OVC). We directly generate the qubit control and readout signals without any analog mixers by using a Xilinx RFSoC evaluation board with QICK firmware~\cite{stefanazzi_qick_2022,ding_experimental_2024} locked to a Rb frequency standard. The signals then pass through rf attenuators, an Eccosorb infrared filter, and a low-pass filter before reaching the sample. The readout signal coming out of the sample goes through a three-wave-mixing Josephson traveling-wave parametric amplifier (TWPA), a cryogenic high-electron-mobility-transistor (HEMT) amplifier, and a room-temperature HEMT amplifier before being digitized by RFSoC. During the first cooldown, the TWPA is not pumped and therefore does not provide any amplification.

\begin{table}[tb!] %
\caption{\textbf{Summary of the measurement results for all four qubits during the first cooldown.} Here, $T_1$ and $T_2^\mathrm{echo}$ are median values. The short $T_1$ and $T_2^\mathrm{echo}$ of qubit Q$_3$ can be explained by the fact that the aluminum film evaporated on top of the resist mask inside its SQUID loop is not removed by the lift-off process.\label{tab:table_results}}
\begin{ruledtabular}
\begin{tabular}{lcccc}
\textrm{Qubit}&
\textrm{$f_\textrm{q}$ (GHz)}&
\textrm{$T_1$ (\textmu s)} & 
\textrm{$T_2^\mathrm{echo}$ (\textmu s)} &
\textrm{$Q$ ($\times$10$^6$)} \\
\colrule
$\mathrm{Q}_1$ (tunable) & 4.047 & 195 & 237 & 5.0 \\
$\mathrm{Q}_2$ (fixed) & 2.890 & N/A\footnote{During the first cooldown, $T_1$ of the qubit $\mathrm{Q}_2$ was measured with such a short waiting time that the qubit did not decay entirely before the next measurement.} & 549 & 9.1 \\
$\mathrm{Q}_3$ (tunable) & 4.593 & 44 & 32 & 1.3\\
$\mathrm{Q}_4$ (fixed) & 3.295 & 154 & 114 & 3.2\\
\end{tabular}
\end{ruledtabular}
\end{table}

\begin{figure*}[tb!]
\includegraphics{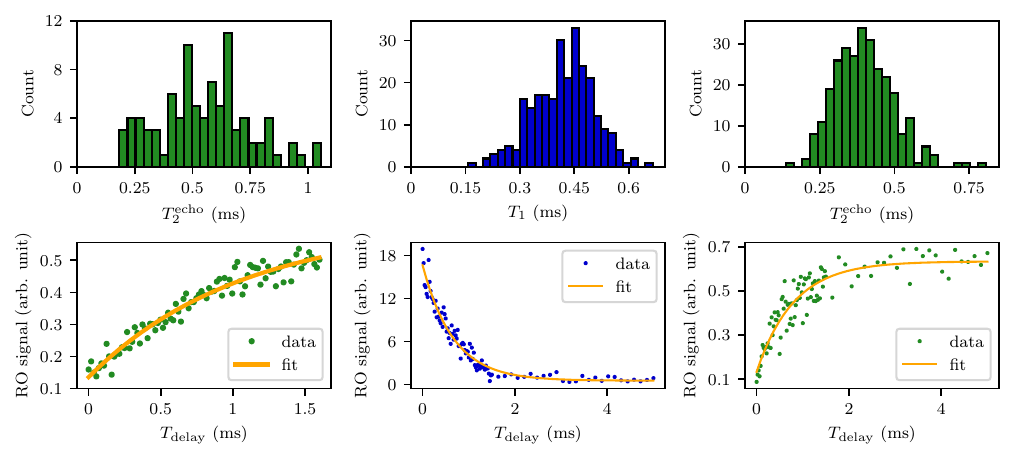} 
\put(-466,218){\small (a)}
\put(-305,218){\small (b)}
\put(-466,109){\small (d)}
\put(-305,109){\small (e)}
\put(-144,218){\small (c)}
\put(-144,109){\small (f)}
\caption{\textbf{Energy relaxation time and echo dephasing time of qubit~$\boldsymbol{\mathrm{Q}_2}$.} (a--c) Distributions of (b) $T_1$ and (a, c) $T_2^\textrm{echo}$ during the (a) first and (b, c) second cooldown. (e) Time trace for the longest measured energy relaxation time during the (e) second cooldown resulting in $T_1 = (666 \pm 33)$~\textmu s. (d, f) Time trace for the longest measured echo dephasing time during the (d) first and (f) second cooldown resulting in  $T_2^\mathrm{echo} = (1057 \pm 138)$~\textmu s and $(806 \pm 78$)~\textmu s, respectively. Panels (b) and (c) include results combined with and without TWPA pump and panel (e) has been obtained with the pump. For other panels, the pump has not been applied.} \label{fig:Histograms_and_best_fits}
\end{figure*}

After the first cooldown, we reconfigure the attenuators as described in the caption of Fig.~\ref{fig:schematic_of_setup} before starting another cooldown. During the second cooldown, we compare measurements with and without a pump signal applied to the TWPA. The 10.5-GHz pump signal passes through a 4--12~GHz circulator, reflects from a 8-GHz low-pass filter, and travels through the circulator again before reaching the TWPA. This configuration allows us to combine the pump and readout signals without the power dissipation of a directional coupler or the impedance mismatch of a diplexer.

\section{Measurement Results} \label{sec:Results}

We measure the fundamental modes of the four readout resonators to have frequencies ranging from $5.85$~GHz to $6.23$~GHz and the fixed-frequency qubits to have qubit frequencies of $f_{\mathrm{Q}_2} = 2.89$~GHz and $f_{\mathrm{Q}_4} = 3.29$~GHz. The flux-tunable qubits have maximum frequencies of $f_{\mathrm{Q}_1} = 4.05$~GHz and $f_{\mathrm{Q}_3} = 4.59$~GHz which we find by sweeping the external magnetic flux applied to the sample. We then follow the standard procedure for measuring the energy relaxation time $T_1$ and echo dephasing time $T_2^\mathrm{echo}$ of each qubit~\cite{krantz_quantum_2019}. The flux-tunable qubits are measured in the qubit sweet-spot, and the frequency dependence of the relaxation and dephasing times is left for future work.

During the first cooldown, we measure that the fixed-frequency qubit $Q_2$ exhibits especially long energy relaxation and echo dephasing times: medians of $T_1 = 502$~\textmu{s} and $T_2^\mathrm{echo}$ of $541$~\textmu{s} and maxima of $T_1 = (765 \pm 0.083)$~\textmu{s} and $T_2^\mathrm{echo} = (1057 \pm 0.138)$~\textmu{s}. However, owing to the short waiting time between each measurement with the qubit $Q_2$, the qubit did not fully relax back to the ground state, resulting in unreliable $T_1$ results for the particular qubit during the first cooldown. The data for these results are shown in Fig.~\ref{fig:Histograms_and_best_fits}(a, b, e, f) in the form of distributions for repeated measurements of $T_1$ and $T_2^\mathrm{echo}$ and time traces for the maximum values. In each $T_1$ time trace, we repeat and average 500 measurements for each value of delay time $T_\mathrm{delay}$ between a $\pi$-pulse and a readout pulse. The number of repetitions is $2000$ for the $T_2^\mathrm{echo}$ measurement. All presented uncertainties correspond to 1$\sigma$ confidence intervals obtained from the fits. 
The measurement results for all four qubits during the first cooldown is summarized in Table~\ref{tab:table_results}.

During the second cooldown, we carry out a more extensive measurement of the qubit $\mathrm{Q}_2$. The qubit frequency is measured to have decreased by $28$~MHz compared to the first cooldown. The readout resonator of this qubit is measured to have a linewidth of 2.63~MHz and a dispersive shift of $2\chi / 2\pi = 1.24$~MHz. Then, we measure the energy relaxation and echo dephasing times with and without the pump signal applied to the TWPA. We fit an exponential function to the obtained data points, each of which is an average of 500 measurements, and compute the relative uncertainties of the fit parameters. We find that the TWPA does not significantly affect the coherence times but improves the certainties of the $T_1$ and $T_2^\mathrm{echo}$ obtained from the fit. These results are presented in Extended Data Fig.~\ref{fig:errors_and_values}. 
We exclude from further analysis the $T_1$ values with relative uncertainties exceeding $5\%$ and $T_2^\mathrm{echo}$ values with relative uncertainties exceeding $10\%$. The resulting data are shown in Fig.~\ref{fig:Histograms_and_best_fits}(c, d, g, h) and Extended Data Fig.~\ref{fig:T1 and T2 results of the second cooldown}. We obtain medians of $T_1 = 425$~\textmu s and $T_2^\mathrm{echo} = 391$~\textmu s and maxima of $T_1 = (666 \pm 33)$~\textmu s and $T_2^\mathrm{echo} = (806 \pm 78)$~\textmu s.
The time trace for the longest measured $T_1$ deviates from an exponential function, likely because of the qubit frequency shifting during the sweep owing to the charge noise.

The data for the coherence time measurements carried out in this section are available in Ref.~\cite{tuokkola_data_2024}. In the Supplementary Note~S1 of the Supplementary Information, we demonstrate the reproducibility of the high-coherence fabrication process by measuring additional transmon qubits in three-dimensional cavity resonators.
 
\section{Discussion} \label{sec:discussion}

In this work, we presented detailed information on our design, fabrication methods, and measurement setup for a high-coherence transmon qubit. One of the four qubits on the measured sample achieved a median energy relaxation time $T_1$ of $425$~\textmu{s} and a median echo dephasing time $T_2^\textrm{echo}$ of $541$~\textmu{s}. These results surpass the previous results for a transmon qubit reported in the literature. However, the $T_1$ and $T_2^\mathrm{echo}$ were significantly shorter in the second cooldown. This may be due to the redistribution of environmental fluctuators and oxidation of the sample surface. 
Our fabrication method and experimental setup can also be applied to other types of superconducting qubits, such as the unimon qubit, to enhance its energy relaxation and dephasing times~\cite{hyyppa_unimon_2022}, and also to large-scale manufacturing as in  Ref.~\cite{van_damme_high-coherence_2024}. 

In conclusion, this result represents a significant step in the development of high-coherence superconducting qubits by approaching the millisecond mark for the energy relaxation and dephasing times of a transmon qubit. It also demonstrates the robustness and reproducibility of the fabrication recipe described in Ref.~\citenum{kono_mechanically_2024} and the effectiveness of the equipment and components listed in Extended Data Tables~\ref{tab:fab equipments} and \ref{tab:measurement equipments}. Detailed reporting of a high-coherence qubit will benefit the research community and accelerate the global efforts on developing quantum sensors, quantum simulators, and quantum computers based on superconducting quantum technologies.



\section*{Data availability}

The data that support the findings of this study are available at \url{https://doi.org/10.5281/zenodo.12819934}.

\begin{acknowledgments}
We acknowledge the Research Council of Finland under its Centre of Excellence Quantum Technology Finland (Grant nos. 336810, 352925, 352934 and 352935) and the Finnish Quantum Flagship, the European Research Council under the Advanced Grant ConceptQ (no. 101053801), the Finnish Quantum Institute InstituteQ, Business Finland through Quantum Technologies Industrial project (no. 41419/31/2020), Horizon Europe programme HORIZON-CL4-2022-QUANTUM-01-SGA through OpenSuperQPlus100 project (no. 101113946), Jane and Aatos Erkko Foundation, QDOC Doctoral Pilot Programme, and the provision of facilities and technical support by Aalto University at the OtaNano-Micronova Nanofabrication Centre.

\end{acknowledgments}

\section*{Competing interests}
M.M. declares that he is a Co-Founder and Shareholder of the quantum-computer company IQM Finland  and of the quantum-algorithm company QMill Oy. J.G and V.V declare that they are Co-Founders and Shareholders of the quantum-hardware company Arctic Instruments Oy. Other authors declare no competing interests.


\putbib
\end{bibunit}

\onecolumngrid

\clearpage

\appendix*

\section{Extended Data Figures and Tables}

\setcounter{figure}{0}
\renewcommand{\figurename}{\textbf{Extended Data Fig.}}
\renewcommand{\theHfigure}{Extended.\thefigure} 

\setcounter{table}{0}
\renewcommand{\tablename}{\textbf{Extended Data Table}}
\renewcommand{\theHtable}{Extended.\thetable}

\begin{table}[thb!]
\caption{\textbf{Comparison of high-coherence transmon qubits in the existing literature and this work.} Qubit frequency $f_q$, median or average energy relaxation time $T_1$, the corresponding quality factor $Q \coloneqq 2\pi f_q T_1$, and median or average echo dephasing time $T_2^\mathrm{echo}$ are shown.}
\label{tab:previous-works}
\begin{ruledtabular}
\begin{tabular}{ccccc}
Ref. & $f_q$ (GHz) & $T_1$ (\textmu s) &$Q$ ($\times$10$^6$) & $T_2^\mathrm{echo}$ (\textmu s) \\
\hline
\citenum{place_new_2021} & 3.73 & 303 & 7.1 & 201 \\
\citenum{wang_towards_2022} & 3.92 & 401 & 9.9 & N/A \\
\citenum{gordon_environmental_2022} & 3.86 & 393 & 9.5 & N/A \\
\citenum{deng_titanium_2023} & 4.24 & 292 & 7.7  & N/A \\
\citenum{sivak_real-time_2023} & 5.92 & 280 & 10.4  & 238 \\
\citenum{biznarova_mitigation_2023} & 3.02 & 270 & 5.1 & 307 \\
\citenum{bal_systematic_2024} & 3.95 & 323 & 8.0 & 286 \\
\citenum{kono_mechanically_2024} & 4.79 & 210 & 6.5 & 290 \\
This work & 2.86 & 425 & 7.7\footnote{Qubit frequency in the relaxation time measurement is 2.89~GHz.}  & 541 \\
\end{tabular}
\end{ruledtabular}
\end{table}

\begin{table}[thb!]
    \caption{\textbf{Equipment used in the fabrication process.}}
    \label{tab:fab equipments}
    \begin{ruledtabular}
    \begin{tabular}{ll}
    Equipment & Description \\ 
    \hline
    Sputtering system & Eclipse Mark IV \\
    Maskless aligner & Heidelberg Instruments MLA150 \\
    \multirow{2}{*}{Plasma processing system} & Oxford Instruments Plasmalab \\
    & 80 Plus \\
    Electron-beam writer & Raith EBPG5200 \\
    Electron-beam evaporator & Plassys MEB700S2-III \\
    Vapor prime oven & YES III \\
    Dicing saw & DISCO DFD6561 \\
    \end{tabular}
    \end{ruledtabular}
\end{table}

\begin{table*}[thb!]
    \centering
        \caption{\textbf{Equipment used in the measurement setup.}}
    \label{tab:measurement equipments}
    \begin{ruledtabular}
    \begin{tabular}{lll} 
    Equipment & Description & Abbreviation\\ 
    \hline
    \hline
    Dilution refrigerator & Bluefors XLD400sl & \\
    \hline
    RFSoC evaluation board & Xilinx Zynq UltraScale+ RFSoC ZCU216 with add-ons CLK104 and XM655 &\\
    \hline
    Frequency standard & Stanford Research Systems FS725 10-MHz Rb Standard & \\
    \hline
    \multirow{5}{*}{Coaxial cables} & Totoku TCF258AA2000/1500 (room temperature) & \\
    & Mini-Circuits 141-72SM+ (room temperature) &\\
    & Bluefors 0.86~mm SCuNi-CuNi semi-rigid (cryogenic input lines) &\\
    & Bluefors 0.86~mm NbTi-NbTi semi-rigid (cryogenic output lines, MXC to 4~K) &\\
    & Bluefors 2.19~mm SCuNi-CuNi semi-rigid (cryogenic output lines, 4~K to RT) &\\
    \hline
    \multirow{2}{*}{HEMT amplifiers} & Low Noise Factory LNF-LNR4\_8F (room temperature) &  HEMT\\
    & Low Noise Factory LNF-LNC4\_8F (cryogenic) & HEMT$^*$\\
    \hline
    \multirow{3}{*}{Low-pass filters} 
    & RLC F-30-8000-R (3-dB point = 8.4~GHz, 60-dB stopband = 10.8--40~GHz) & LPF\\
    & Mini-Circuits VLF-5850+ (3-dB point = 6.54~GHz) & LPF$^*$ \\
    & Aivon Therma uD25 pairwise RC-filter (1.5~k\textOmega\ per wire, 3-dB point = 1~kHz) & LPF$^\dagger$ \\
    \hline
    \multirow{2}{*}{Bandpass filters} 
    & Mini-Circuits VBFZ-3590+ (2-dB passband = 3--4.3~GHz) & VBFZ\\
    & Mini-Circuits VBFZ-6260-S+ (2-dB passband = 5.6--7~GHz) & VBFZ$^*$\\
    \hline
    Highpass filter & Mini-Circuits VHF-8400+ (20-dB stopband = DC--7.8~GHz) & HPF\\
    \hline
    Eccosorb filter & Kawashima Manufacturing 3.5mm Filter Connector L9.6 (0.32~dB/GHz) & Ecco.\\
    \hline
    \multirow{3}{*}{Attenuator} & Bluefors cryogenic attenuators & $X$ dB \\
    & XMA 2082-6418-20-CRYO & 20~dB$^*$ \\
    & Mini-Circuits BW-S10-2W263+ (room temperature) & 10~dB$^\dagger$ \\
    \hline
    \multirow{2}{*}{Isolators} & Low Noise Factory LNF-ISISC4\_8A (4--8~GHz) &\\
    & Low Noise Factory LNF-ISISC4\_12A (4--12~GHz) &\\
    \hline
    Circulator & Low Noise Factory LNF-CIC4\_12A (4--12~GHz)&\\
    \hline
    TWPA & Supplied by VTT (Technical Research Center of Finland Ltd.) & TWPA\\
    \hline
    TWPA pump generator & Valon 5015 Frequency Synthesizer &\\
    \hline
    Voltage source & Stanford Research Systems SIM928 Isolated Voltage Source & V \\
    \hline
    Sample holder & QDevil QCage.24 with QCage Magnetic Shielding & QCage.24\\
    \end{tabular}
    \end{ruledtabular}
\end{table*}

\clearpage

\begin{figure*}[h!]
\centering
\subfigure{
    \begin{minipage}[b]{188.25pt}
    \centering
    \includegraphics[width=\textwidth]{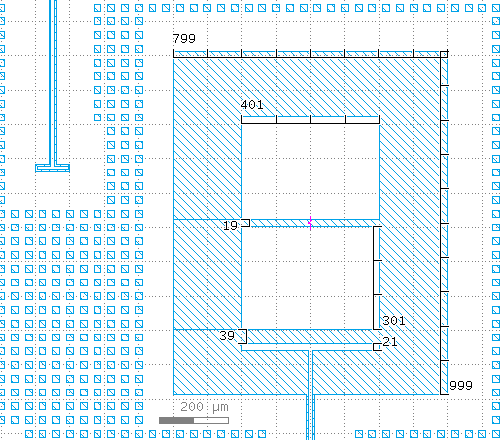}
    \put(-190,170){\small \colorbox{white}{(a)}}
    \end{minipage}
}
\subfigure{
    \begin{minipage}[b]{83.25pt}
    \centering
    \includegraphics[width=\textwidth]{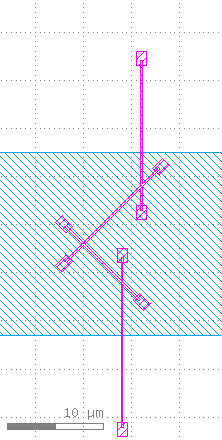}
    \put(-85,170){\small \colorbox{white}{(b)}}
    \end{minipage}
}
\subfigure{
    \begin{minipage}[b]{145.125pt}
    \centering
    \includegraphics[width=\textwidth]{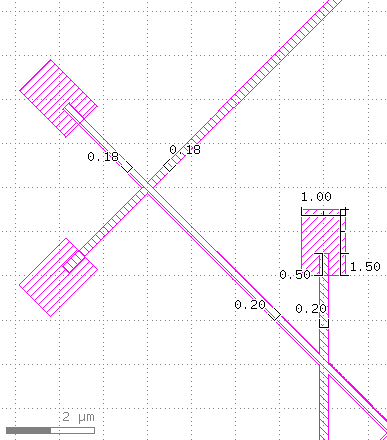}
    \put(-145,170){\small \colorbox{white}{(c)}}
    \end{minipage}
}
\caption{\textbf{Lithographic pattern of qubit~$\boldsymbol{\mathrm{Q}_2}$.} (a)~Overview of our transmon qubit with its drive line (top left) and coupling capacitor of the readout resonator (bottom). The Nb film in the shaded region is removed by photolithography and dry etching. (b)~Close-up of the pattern for the electron-beam lithography of the Josephson junction (diagonal strips) and the bandages for galvanically connecting it to the Nb capacitor (vertical strips). The rectangular parts at the tips of each strip are exposed at a lower dose than the other exposed regions and becomes an undercut in the two-layer resist stack. (c)~Close-up of the Josephson junction. The dimensions of the Josephson junction are 180~nm by 180~nm in the mask pattern.\label{fig:gds_images}}
\end{figure*}

\begin{figure*}[h!]
\centering
\subfigure{
    \begin{minipage}[b]{230pt}
    \centering
    \includegraphics[width=\textwidth]{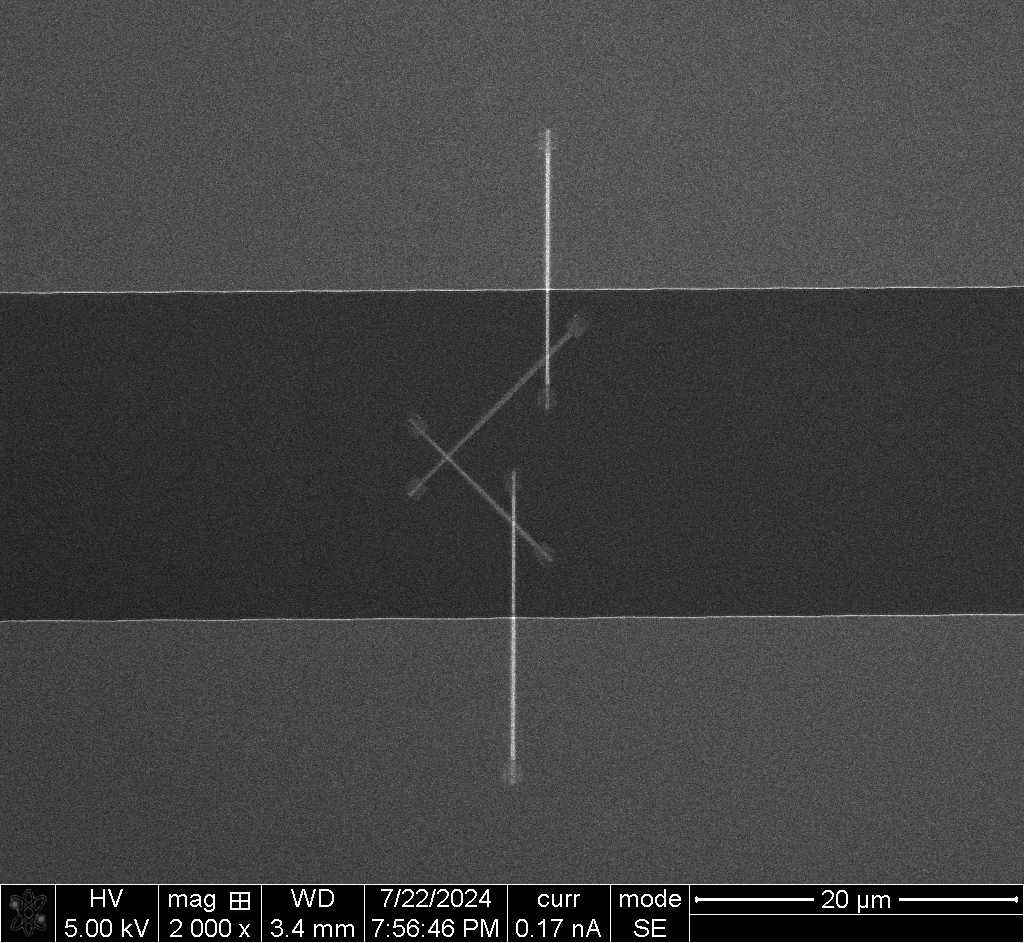}
    \put(-228,200){\small \colorbox{white}{(a)}}
    \end{minipage}
}
\subfigure{
    \begin{minipage}[b]{230pt}
    \centering
    \includegraphics[width=\textwidth]{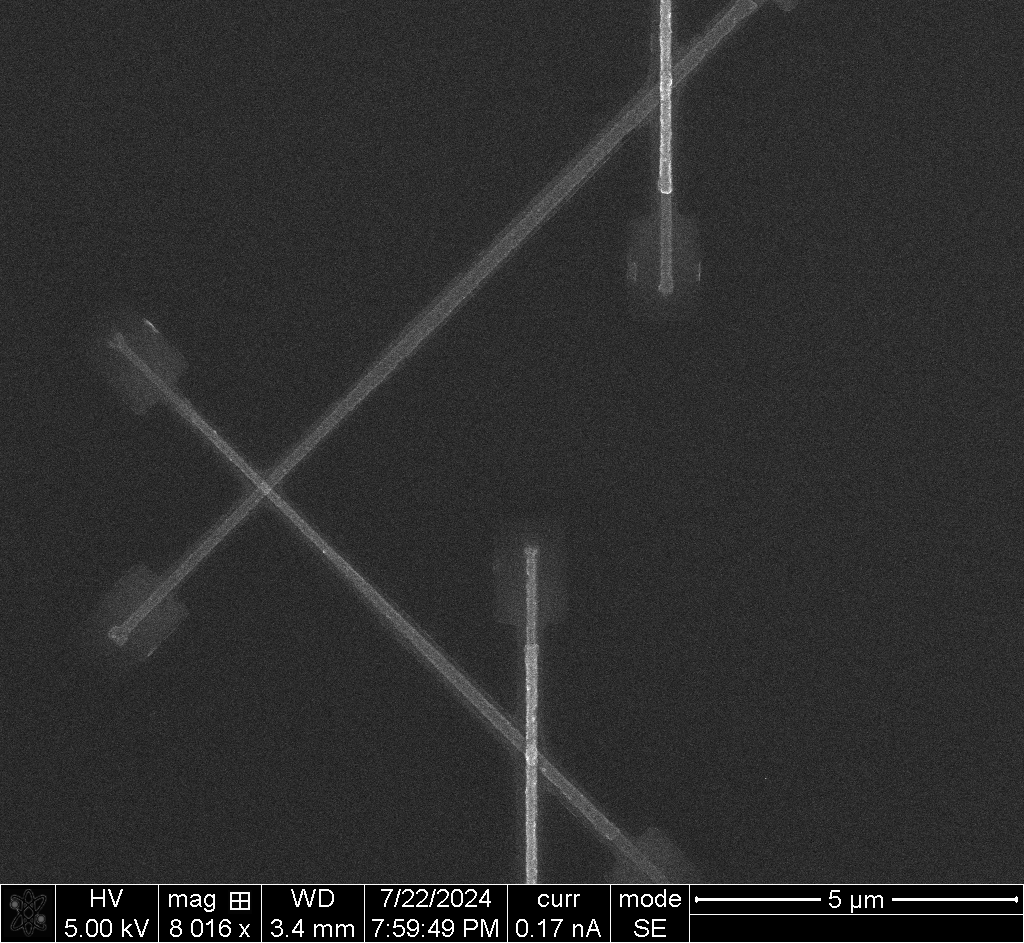}
    \put(-228,200){\small \colorbox{white}{(b)}}
    \end{minipage}
}
\subfigure{
    \begin{minipage}[b]{230pt}
    \centering
    \includegraphics[width=\textwidth]{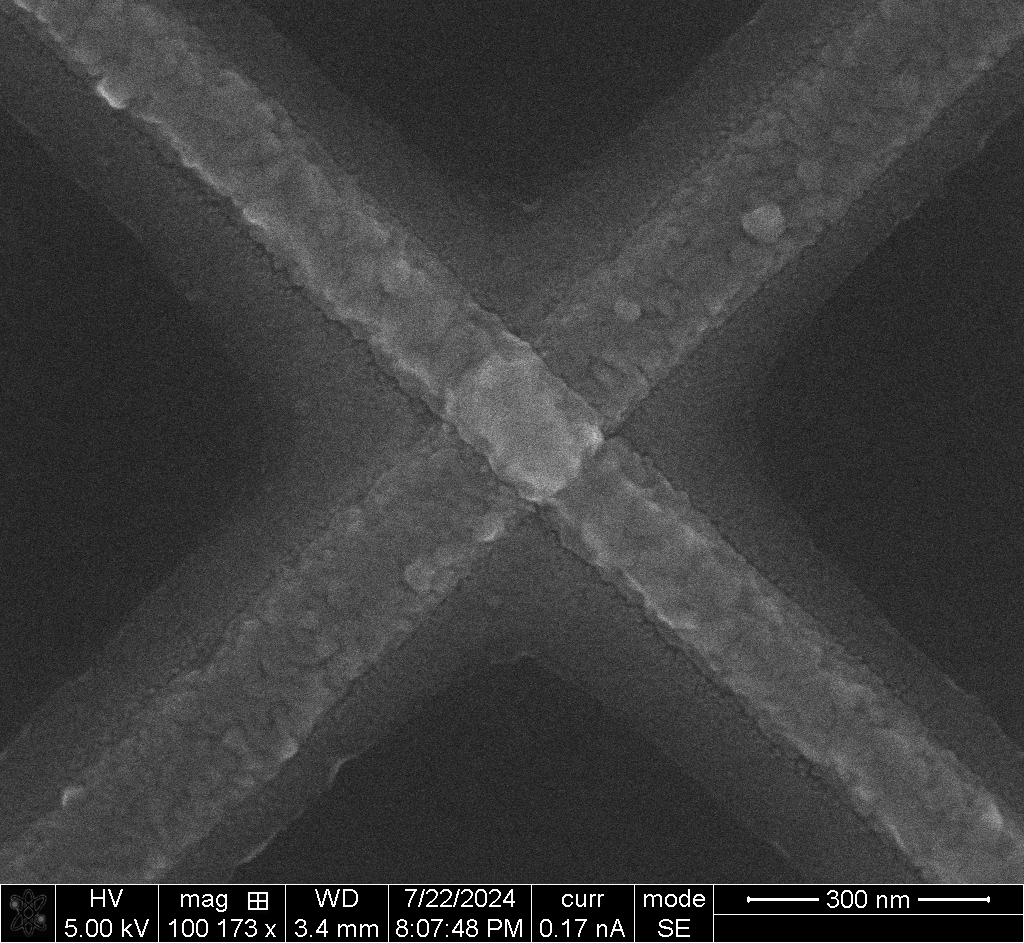}
    \put(-228,200){\small \colorbox{white}{(c)}}
    \end{minipage}
}
\subfigure{
    \begin{minipage}[b]{230pt}
    \centering
    \includegraphics[width=\textwidth]{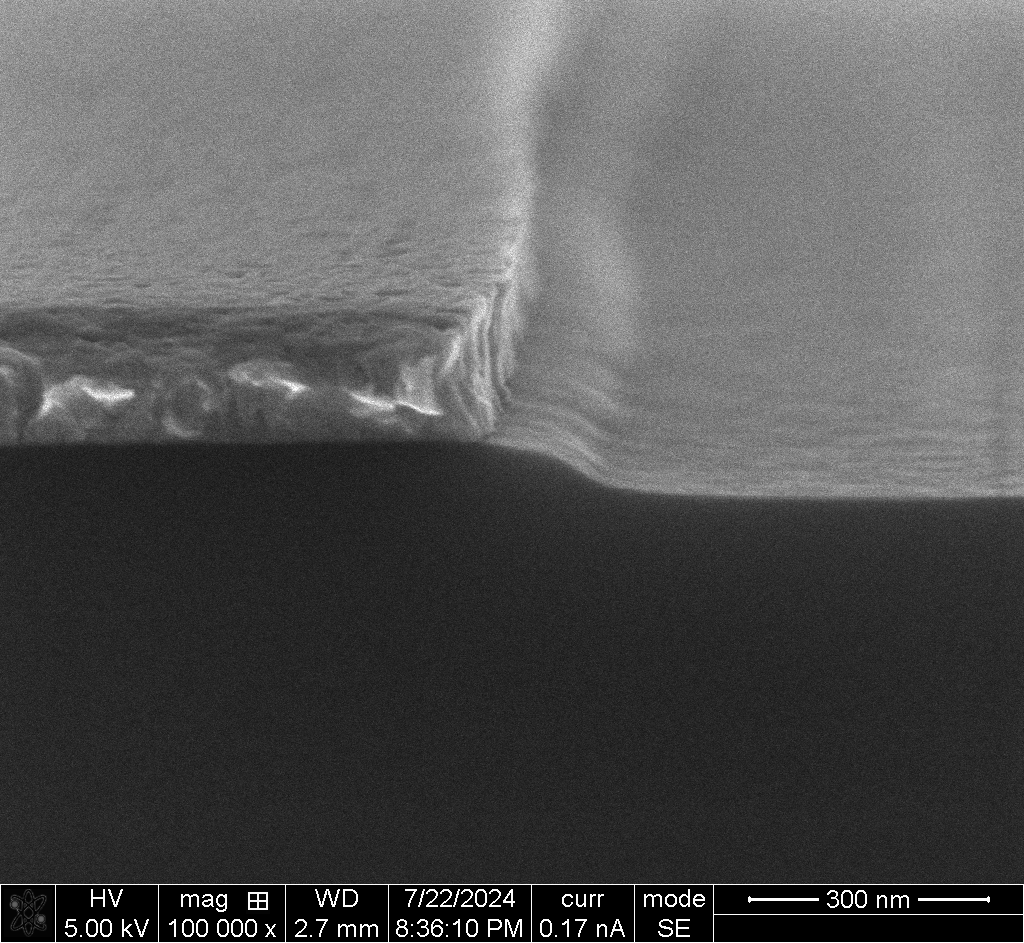}
    \put(-228,200){\small \colorbox{white}{(d)}}
    \end{minipage}
}
\caption{\textbf{Scanning-electron-microscope (SEM) images of a chip processed in the same fabrication run as the sample measured in this work.} (a)~Josephson junction (diagonal strips) and the bandages (vertical strips) for galvanically connecting it to the Nb capacitor (top and bottom gray regions). (b)~Close-up of the Josephson junction. (c)~Further close-up of the Josephson junction. The junction shown here, which is fabricated using a 140~nm by 140~nm mask pattern, is measured to be approximately 170~nm by 140~nm. This implies that the junction of qubit~$\mathrm{Q}_2$ is likely approximately 210~nm by 180~nm. (d)~Etch profile of the Nb film (left) on the Si substrate (bottom) obtained by cleaving the chip.\label{fig:sem_images}}
\end{figure*}

\begin{figure}[tb!]
\includegraphics{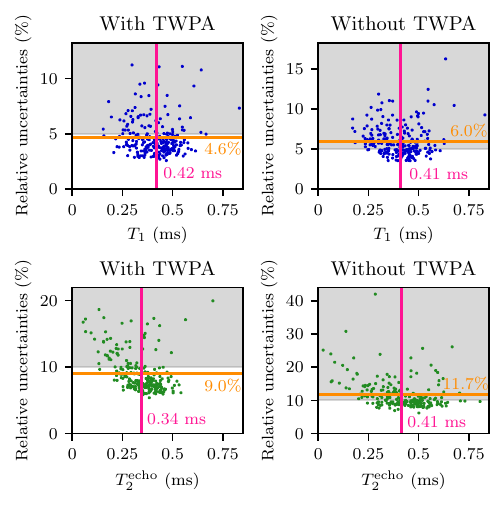} 
\put(-210,229){\small (a)}
\put(-97,229){\small (b)}
\put(-210,112){\small (c)}
\put(-97,112){\small (d)}
\caption{\textbf{Energy relaxation and echo dephasing times and their relative uncertainties for qubit $\boldsymbol{\mathrm{Q}_2}$ measured during the second cooldown.} (a) $T_1$ measured while pumping the TWPA. (b) $T_1$ measured without pumping the TWPA. (c) $T_2^\textrm{echo}$ measured while pumping the TWPA. (d) $T_2^\textrm{echo}$ measured without pumping the TWPA. The orange horizontal lines and the pink vertical lines represent the average uncertainties and the median values of $T_1$ or $T_2^\textrm{echo}$, respectively. The values in the grey area are excluded from further analysis.\label{fig:errors_and_values}}
\end{figure}

\begin{figure}[thb!]
\includegraphics{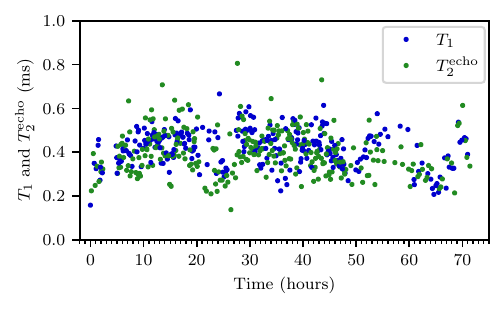} 
\caption{\textbf{Stability of the energy relaxation time $T_1$ and echo dephasing time $T_2^\mathrm{echo}$ measured during the second cooldown}. Energy relaxation and echo dephasing time as function of time over a $70$-h period. \label{fig:T1 and T2 results of the second cooldown}}
\end{figure}

\clearpage

\pagenumbering{roman} 
\setcounter{page}{1} 
\section*{Supplementary Information for "Methods to achieve near-millisecond energy relaxation and dephasing times for a superconducting transmon qubit"} \label{Supplementary section S1}
\renewcommand{\thefigure}{S\arabic{figure}}
\renewcommand{\thetable}{S\arabic{table}}
\renewcommand{\figurename}{\textbf{Supplementary Fig.}} 
\setcounter{figure}{0} 
\renewcommand{\tablename}{\textbf{Supplementary Table}} 
\setcounter{table}{0} 

\begin{bibunit}

\section*{Supplementary Note S1: High-coherence transmon qubit in a 3D cavity } \label{Supplementary S1}

In this supplementary section, we present high-coherence transmon qubits coupled to three-dimensional (3D) cavity resonators to demonstrate the reproducibility of the fabrication process described in the main article. We measure two additional qubits $\mathrm{Q}_{\text{S}1}$ and $\mathrm{Q}_{\text{S}2}$ fabricated with an almost identical recipe as described in Sec.~\ref{sec:Fabrication process}. 
The only differences are the HF etching time which is 5~min instead of 10~min, and the junction oxidation pressure which is 1.5~mbar instead of 1.2~mbar. The qubits have an identical design, which is shown in Supplementary Fig.~\ref{fig:transmon_3d_gds} and is available on Zenodo \cite{tuokkola_data_2024}. 

\begin{figure}[b!]
    \centering
    \includegraphics[height=120pt]{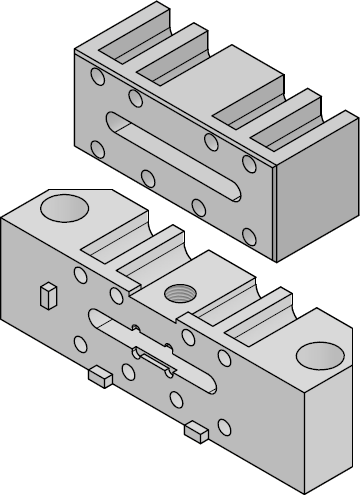}
    \qquad \qquad \qquad
    \includegraphics[height=120pt]{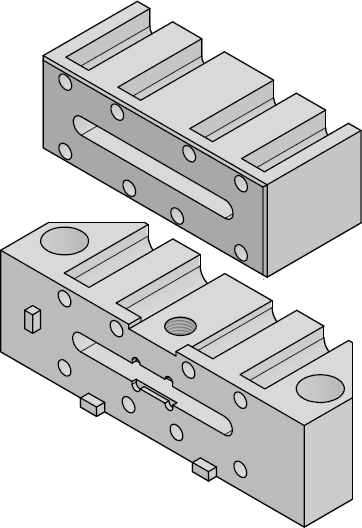}
    \put(-235,120){\small (a)}
    \put(-100,120){\small (b)}
    \caption{{\bf Supplementary cavity design.} (a, b) Design illustrations of the 3D aluminum cavities (a) C$_{16}$ and (b) C$_{20}$ used in the experiments of this supplementary note. See full design file in Zenodo.}
    \label{fig:cavities}
\end{figure}

Qubits $\mathrm{Q}_{\text{S}1}$ and $\mathrm{Q}_{\text{S}2}$ are measured in 3D cavity resonators, C$_{20}$ and C$_{16}$, respectively.
Supplementary Fig.~\ref{fig:cavities} presents illustrations of the cavities, which are made of 1050 aluminum alloy, and the design is available on Zenodo \cite{tuokkola_data_2024}.

\begin{figure}[b!]
    \centering
    \includegraphics[height=180pt]{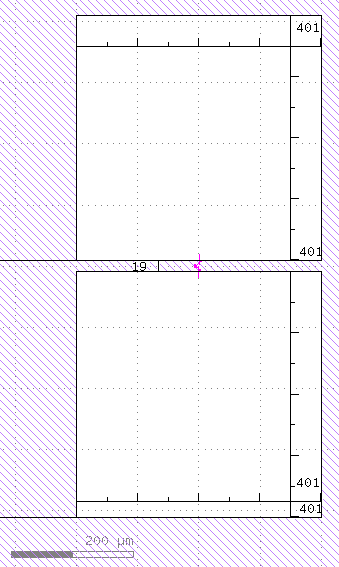}
    \includegraphics[height=180pt]{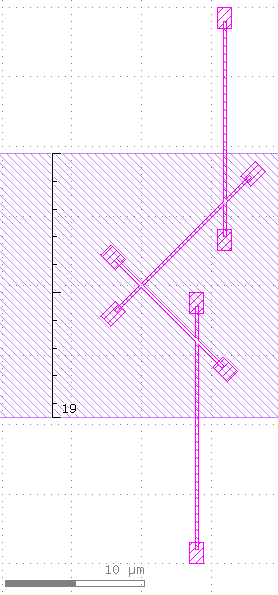}
    \includegraphics[height=180pt]{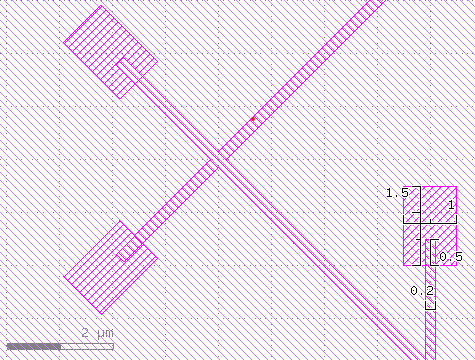}
    \put(-435,184){\small (a)}
    \put(-324,184){\small (b)}
    \put(-235,184){\small (c)}
    \caption{{\bf Supplementary sample design.} (a) Design of the transmon samples used in the measurements of this Supplementary Note~S1 and (b, c) magnifications in the vicinity of the Josephson junction. See full design file in Zenodo.}
    \label{fig:transmon_3d_gds}
\end{figure}

The qubits are measured in the same dilution refrigerator and with the same evaluation board as in the main article. Unlike in the main article, in this measurement setup, the readout signal and the qubit drive signal are combined into a single line by using a directional coupler, and we measure the reflection from the cavity instead of the transmission. The measurement setup is presented in the schematic diagram of Supplementary Fig.~\ref{fig: measurement setup 3D}, and the equipment is listed in Tables~\ref{tab:measurement equipments}~and~\ref{tab:measurement equipment 3d}. The cavity is protected from electromagnetic fields with a custom-designed four-layer magnetic shield. From the innermost to the outermost layer, the shield consists of 1.5-mm-thick gold-plated copper (C10100), 1.5-mm-thick aluminum 1050, and two layers of 1.5-mm-thick mumetal (Hymu80). 

\begin{figure}[tb!]
    \centering
    \includegraphics{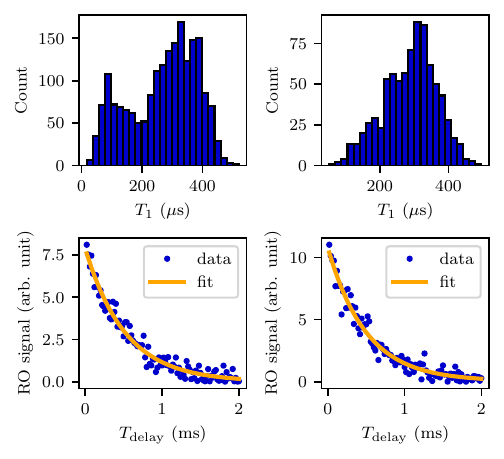}
    \put(-210,220){\small (a)}
\put(-94,220){\small (b)}
\put(-210,113){\small (c)}
\put(-94,113){\small (d)}
    \caption{\textbf{Energy relaxation times of the additional qubits~$\boldsymbol{\mathrm{Q}_{\text{S}1}}$~and~$\boldsymbol{\mathrm{Q}_{\text{S}2}}$.} (a, b) 
    Distributions of the measured energy relaxation times of the qubits (a) ~$\mathrm{Q}_{\text{S}1}$ over 48 hours and (b) ~$\mathrm{Q}_{\text{S}2}$ over 40 hours.
    (c, d) Time traces for the longest measured relaxation times of (c) the qubit~$\mathrm{Q}_{\text{S}1}$ with $\mathrm{T}_1 = (521 \pm 27)$~\textmu{s} and (d) the qubit~$\mathrm{Q}_{\text{S}2}$ with $\mathrm{T}_1 = (496 \pm 26)$~\textmu{s}. 
    }    \label{fig:3d_transmon_results}
\end{figure}

We measure that the cavities C$_{20}$ and C$_{16}$ have resonance frequencies of 5.55~GHz and 6.32~GHz, respectively. The frequency of qubit~$\mathrm{Q}_{\text{S}1}$ is measured to be 2.9755~GHz and with an anharmonicity of $-226$~MHz. These parameters are 2.694~GHz and $-230$~MHz for the qubit~$\mathrm{Q}_{\text{S}2}$, respectively. We measure the qubit~$\mathrm{Q}_{\text{S}1}$ to have a median energy relaxation time $T_1$ of $270$~\textmu{s} with the longest measured $T_1$ of $(521 \pm 27)$~\textmu{s}. For the qubit~$\mathrm{Q}_{\text{S}2}$, the median $T_1$ is measured to be $288$~\textmu{s} and the longest $T_1$ to be $(496 \pm 26)$~\textmu{s}. Supplementary Fig.~\ref{fig:3d_transmon_results} presents distributions of the measured relaxation times and the time traces of the longest measured relaxation times. The data used to generate the figures are published on Zenodo \cite{tuokkola_data_2024}.

\begin{figure}[b!]
\includegraphics{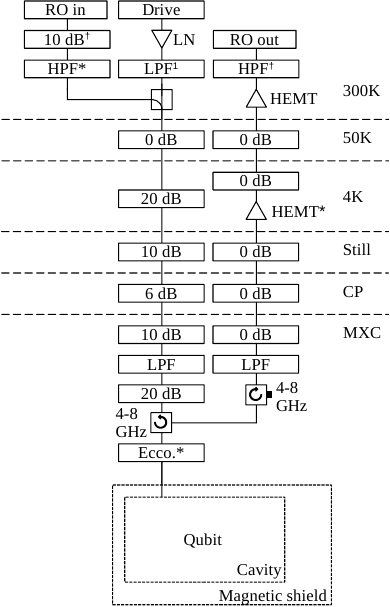}
\caption{\textbf{Schematic of the measurement setup used in measurements of the qubits in cavities.} See Extended Data Table~\ref{tab:measurement 
equipments} and Supplementary Table~\ref{tab:measurement equipment 3d} for a description of each component. }
\label{fig: measurement setup 3D}
\end{figure}

\begin{table*}[b!]
    \centering
    \caption{\textbf{Equipment used in the measurement setup of the supplementary note.}}
    \label{tab:measurement equipment 3d}
    \begin{ruledtabular}
    \begin{tabular}{lll} 
    Equipment & Description & Abbreviation\\ 
    \hline
    Low-pass filters 
    & Mini-Circuits VLF-3400+ (3-dB point = 3.8~GHz) & LPF$^1$\\
    \hline
    Low-noise amplifier 
    & Mini-Circuits ZX60-123LN-S+ & LN\\
        \hline
    Directional coupler & Mini-Circuits ZUDC20-02183-S+&\\
    \hline
     \multirow{2}{*}{High-pass filter} & Mini-Circuits VHF-3100+ (20-dB stopband = DC--2.5~GHz) & HPF$^*$\\
      & Mini-Circuits VHF-4400+ (20-dB stopband = DC--3.85~GHz) & HPF$^\dagger$\\
    \hline
    Eccosorb filter & Kawashima Manufacturing 3.5mm Filter Connector L5 (0.32~dB/GHz) & Ecco.$^*$\\
    \hline
    Circulator & Low Noise Factory LNF-CIC4\_8A (4--8~GHz)&\\
    \hline
    \end{tabular}
    \end{ruledtabular}
\end{table*}

\FloatBarrier
\putbib
\end{bibunit}

\end{document}